# Energy band of graphene ribbons under the tensile force

Yong Wei, Guo-Ping Tong*, and Sheng Li

Institute of Theoretical Physics, Zhejiang Normal University, Jinhua 321004, Zhejiang, China

According to the tight-binding approximation, we investigate the electronic structures of graphene ribbons with zigzag shaped edges (ZGRs) and armchair shaped edges (AGRs) drawn by the tensile force, and obtain the analytic relations between the energy bands of π-electrons in ZGR, AGR and the tensile force based on only considering the nearest-neighbor interaction and the hydrogen-like atomic wave function is considered as π-electron wave function. Importantly, we find the tensile force can open an energy gap at the *K* point for ZGR and AGR, and the force perpendicular to the zigzag edges can open energy gap more easily besides the gap values of ZGR and AGR at the *K* point both increase as the tensile force increases.

Graphene, a single atomic layer of graphite, has attracted much attention recently,[1-3] because it not only has been produced successfully in experiment[4] and observed by an ordinary optical microscope[5,6], but also has a number of unusual transport properties[7-9] and exists the Klein paradox[10], weak localization[11], etc. Moreover, the graphene has many potential applications.[2] The electronic structure of graphene has been investigated by Wallace since 1947.[12] But the energy dispersion relations of the $\pi$-electrons of graphene were given very well by Saito et al. using the tight-binding approximation although only considering the nearest-neighbor interaction.[13] Then S.Reich et al. investigated the energy band of graphene using the same method including up to third-nearest neighbors in 2002.[14] Nowadays, A. H. Castro Neto et al. give a description of conclusion for the electronic properties of graphene.[2]

Because of some significant properties of graphene that can be future electronic devices, graphene nanoribbon (GNR), a strip of graphene of nanometers in width, with zigzag shaped edges or armchair shaped edges on both sides, and graphene ribbon that the width is far more than nanometer, have also investigated by many investigators.[3,15-21] Whether the graphene nanoribbon or the graphene ribbon, it must be effected by the tensile force when it is been electronic devices, but it is not clear that the relations between the electronic structure of graphene and the tensile force, which is solved by this paper.

In this paper, we build a simple model: the graphene ribbon with zigzag shaped edges or armchair shaped edges on both sides is drawn by the tensile force perpendicular to two graphene ribbon edges and parallel to the graphene ribbon plane (see Figs.1(a) and (b)), the widths of graphene ribbons are so wide that the periodic boundary condition is satisfied and the approximation of an infinite plane is used in calculating electronic structures of graphene ribbons, besides the interaction between carbon atoms satisfies the Huke'law. Then we investigate the relations between the π-electron bands of the graphene ribbons with zigzag shaped edges (ZGR), armchair shaped edges (AGR) and the tensile force based on the tight-binding approximation.

We first investigate the ZGR drawn by the tensile force. Similar to the method of Ref.13 using the tight-binding approximation and only considering the nearest-neighbor interaction, we obtain the matrix Hamiltonian $H_{AA}$ as

$$H_{AA}(\vec{r}) = \frac{1}{N}\sum_{\vec{R}} e^{i\vec{k}\cdot(\vec{R}-\vec{R})} \left\langle \varphi_A\left(\vec{r}-\vec{R}\right)\middle| H \middle| \varphi_A\left(\vec{r}-\vec{R}\right)\right\rangle$$

$$= \frac{1}{N}\sum_{\vec{R}} \varepsilon_{2p} = \varepsilon_{2p}, \qquad (1)$$

where $N$ is the unit cell number of ZGR, $\vec{R}$ is the site vector of carbon atom $A$, $\vec{k}$ is the wave vector, $H$ is a one-electron Hamiltonian, $\varphi$ can be considered as $2p_z$-orbital wave function of carbon atoms. Here $\varepsilon_{2p} = \left\langle \varphi_A\left(\vec{r}-\vec{R}\right)\middle| H \middle| \varphi_A\left(\vec{r}-\vec{R}\right)\right\rangle$ is different from the atomic energy value for the free carbon atom because the Hamiltonian $H$ contains all ZGR potential and changes as the tensile force changes, so $\varepsilon_{2p}$ is a variable as the tensile force. Similarly, $H_{BB} = \varepsilon_{2p}$ for the same order of approximation.

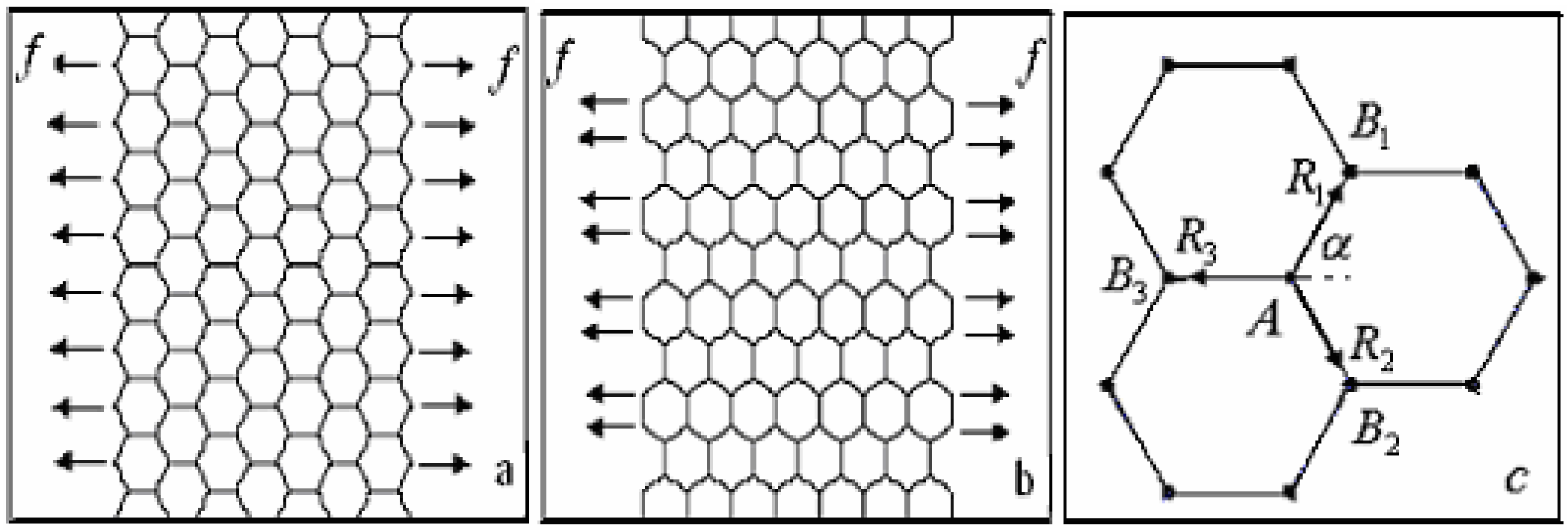

**FIG. 1.** (a).The graph of graphene ribbon with zigzag shaped edges (ZGR) drawn by the tensile force. (b).The graph of graphene ribbon with armchair shaped edges (AGR) drawn by the tensile force. (c).The graph of showing site vectors $\vec{R}_1$, $\vec{R}_2$, $\vec{R}_3$ and half-bond angle $\alpha$.

We also obtain the matrix Hamiltonian $H_{AB}$ for only considering the nearest-neighbor interaction as

$$H_{AB} = \frac{1}{N}\sum_{\vec{R}} \Big\{ e^{i\vec{k}\cdot\vec{R}_1} \left\langle \varphi_A\left(\vec{r}-\vec{R}\right)\middle| H \middle| \varphi_{B_1}\left(\vec{r}-\vec{R}-\vec{R}_1\right)\right\rangle$$

$$+ e^{i\vec{k}\cdot\vec{R}_2} \left\langle \varphi_A\left(\vec{r}-\vec{R}\right)\middle| H \middle| \varphi_{B_2}\left(\vec{r}-\vec{R}-\vec{R}_2\right)\right\rangle$$

$$+e^{i\vec{k}\cdot\vec{R}_3}\left\langle \varphi_A\left(\vec{r}-\vec{R}\right)\middle| H \middle| \varphi_{B_3}\left(\vec{r}-\vec{R}-\vec{R}_3\right)\right\rangle\Big\}$$

$$=\gamma_1 e^{i\vec{k}\cdot\vec{R}_1}+\gamma_2 e^{i\vec{k}\cdot\vec{R}_2}+\gamma_3 e^{i\vec{k}\cdot\vec{R}_3}, \tag{2}$$

where $\vec{R}_1, \vec{R}_2$ and $\vec{R}_3$ are the positions of the nearest-neighbor atom $B$ relative to atom $A$ (see Fig.1(c)), the transfer integrals $\gamma_1$, $\gamma_2$ and $\gamma_3$ are denoted by

$$\gamma_1=\left\langle \varphi_A\left(\vec{r}-\vec{R}\right)\middle| H \middle| \varphi_{B_1}\left(\vec{r}-\vec{R}-\vec{R}_1\right)\right\rangle,$$

$$\gamma_2=\left\langle \varphi_A\left(\vec{r}-\vec{R}\right)\middle| H \middle| \varphi_{B_2}\left(\vec{r}-\vec{R}-\vec{R}_2\right)\right\rangle,$$

$$\gamma_3=\left\langle \varphi_A\left(\vec{r}-\vec{R}\right)\middle| H \middle| \varphi_{B_3}\left(\vec{r}-\vec{R}-\vec{R}_3\right)\right\rangle, \tag{3}$$

where $\gamma_1=\gamma_2$ due to $R_1=R_2$ according to the symmetry of ZGR. The matrix element $H_{BA}$ can be obtained by the Hermitian conjugation relation $H_{BA}=H_{AB}^*$. Using the similar method used for $H_{AA}$ or $H_{AB}$, we obtain the overlap matrix $S_{AA}=S_{BB}=1$ as assuming that the atomic wave function is normalized and also derive $S_{AB}=s_1 e^{i\vec{k}\cdot\vec{R}_1}+s_2 e^{i\vec{k}\cdot\vec{R}_2}+s_3 e^{i\vec{k}\cdot\vec{R}_3}$, where the overlap integrals: $s_1$, $s_2$ and $s_3$ between the nearest $A$ and $B$ atoms are denoted by

$$s_1=\left\langle \varphi_A\left(\vec{r}-\vec{R}\right)\middle| \varphi_{B_1}\left(\vec{r}-\vec{R}-\vec{R}_1\right)\right\rangle,$$

$$s_2=\left\langle \varphi_A\left(\vec{r}-\vec{R}\right)\middle| \varphi_{B_2}\left(\vec{r}-\vec{R}-\vec{R}_2\right)\right\rangle,$$

$$s_3=\left\langle \varphi_A\left(\vec{r}-\vec{R}\right)\middle| \varphi_{B_3}\left(\vec{r}-\vec{R}-\vec{R}_3\right)\right\rangle, \tag{4}$$

we can get $s_1=s_2$ because of the symmetry and obtain $S_{BA}=S_{AB}^*$ though the Hermitian conjugation relation. We solve the secular equation $\det\left[H-ES\right]=0$ and obtain the energy dispersion relations of the $\pi$-electrons of ZGR as

$$E^{\pm}\left(\vec{k}\right)=\frac{\varepsilon_{2p}-\gamma_3 s_3-O\left(\vec{k}\right)\pm\sqrt{\left(\gamma_3-\varepsilon_{2p}s_3\right)^2+P\left(\vec{k}\right)}}{1-s_3^{\,2}-Q\left(\vec{k}\right)}, \tag{5}$$

where the $O\left(\vec{k}\right)$, $P\left(\vec{k}\right)$ and $Q\left(\vec{k}\right)$ are denoted by

$$O\left(\vec{k}\right)=2\gamma_1 s_1\left[1+\cos\left(\vec{k}\cdot\vec{R}_1-\vec{k}\cdot\vec{R}_2\right)\right]$$
$$+\left(\gamma_1 s_3+\gamma_3 s_1\right)\left[\cos\left(\vec{k}\cdot\vec{R}_1-\vec{k}\cdot\vec{R}_3\right)+\cos\left(\vec{k}\cdot\vec{R}_2-\vec{k}\cdot\vec{R}_3\right)\right], \tag{6}$$

$$P\left(\vec{k}\right)=2\left(\gamma_1-\varepsilon_{2p}s_1\right)\left(\gamma_3-\varepsilon_{2p}s_3\right)\left[\cos\left(\vec{k}\cdot\vec{R}_1-\vec{k}\cdot\vec{R}_3\right)+\cos\left(\vec{k}\cdot\vec{R}_2-\vec{k}\cdot\vec{R}_3\right)\right]$$
$$+2\left(\gamma_1-\varepsilon_{2p}s_1\right)^2\left[1+\cos\left(\vec{k}\cdot\vec{R}_1-\vec{k}\cdot\vec{R}_2\right)\right]$$

$$-\left(\gamma_1 s_3-\gamma_3 s_1\right)^2\left[\sin\left(\vec{k}\cdot\vec{R}_1-\vec{k}\cdot\vec{R}_3\right)+\sin\left(\vec{k}\cdot\vec{R}_2-\vec{k}\cdot\vec{R}_3\right)\right]^2, \tag{7}$$

$$Q\left(\vec{k}\right)=2s_1^{\ 2}\left[1+\cos\left(\vec{k}\cdot\vec{R}_1-\vec{k}\cdot\vec{R}_2\right)\right]$$
$$-2s_1 s_3\left[\cos\left(\vec{k}\cdot\vec{R}_1-\vec{k}\cdot\vec{R}_3\right)+\cos\left(\vec{k}\cdot\vec{R}_2-\vec{k}\cdot\vec{R}_3\right)\right], \tag{8}$$

$E^+\left(\vec{k}\right)$ and $E^-\left(\vec{k}\right)$ are called bonding $\pi$ and antibonding $\pi^*$ energy bands. Now we explore the relations between the parameters of Eq. (5) and the tensile force in order to obtain the analytic expressions of $\pi$-electron energy band with the tensile force. According to the geometric knowledge, we can easy to obtain the lattice vectors $\vec{a}_1$ , $\vec{a}_2$ and the reciprocal lattice vectors $\vec{b}_1$ , $\vec{b}_2$ of ZGR as

$$\vec{a}_1=\left(R_1\cos\alpha+R_3\,,R_1\sin\alpha\right),$$
$$\vec{a}_2=\left(R_1\cos\alpha+R_3\,,-R_1\sin\alpha\right), \tag{9}$$

$$\vec{b}_1=\left(\frac{\pi}{R_1\cos\alpha+R_3},\frac{\pi}{R_1\sin\alpha}\right),$$
$$\vec{b}_2=\left(\frac{\pi}{R_1\cos\alpha+R_3},-\frac{\pi}{R_1\sin\alpha}\right), \tag{10}$$

where the relative positions $R_1, R_2, R_3$ and the half-bond angle $\alpha$ (shown in Fig.1(c)) are given due to the mechanical and geometric knowledge by

$$R_1=a_{c-c}+\Delta R_1=\frac{\sqrt{3}a_{c-c}}{2\sin\left(\frac{\pi}{3}+\frac{\sqrt{3}}{6}\left[(1+f_N)-\sqrt{1+14f_N+f_N^2}\right]\right)},$$

$$R_2=R_1,$$

$$R_3=a_{c-c}+\Delta R_3=a_{c-c}(1+f_N),$$

$$\alpha=\frac{\pi}{3}+\frac{\sqrt{3}}{6}\left[(1+f_N)-\sqrt{1+14f_N+f_N^2}\right], \tag{11}$$

where $a_{c-c}$=0.142 $nm$ is the nearest-neighbor distance between two carbon atoms without the tensile force, $f_N$ is a dimensionless value of $f$ that the tensile force shares to a little component force of each carbon atom and $f_N = f/(k_0 a_{c-c})$ , $k_0$ is the elastic constant between two carbon atoms. According to Eq. (10), we can also obtain the positions of the Dirac points $K$ and $K'$ in momentum space as

$$\vec{K}=\left(\frac{\pi}{R_1\cos\alpha+R_3},\frac{\pi}{2R_1\sin\alpha}-\frac{\pi R_1\sin\alpha}{2(R_1\cos\alpha+R_3)^2}\right),$$

$$\vec{K}'=\left(\frac{\pi}{R_1\cos\alpha+R_3},\frac{\pi R_1\sin\alpha}{2(R_1\cos\alpha+R_3)^2}-\frac{\pi}{2R_1\sin\alpha}\right). \tag{12}$$

Based on Eqs. (5), (6), (7), (8), (10) and (11), if we can also obtain the relations between the transfer integrals, the overlap integrals and the tensile force, we will obtain the analytic relations between the energy band of the $\pi$-electrons of ZGR and the tensile force $f_N$ in all Brillouin zone. For one-electron Hamiltonian $H$, if we only consider the nearest-neighbor interaction, we can write $H$ approximately as

$$H=-\frac{1}{2}\nabla^2+V_A\left(\vec{r}-\vec{R}\right)+V_{B_1}\left(\vec{r}-\vec{R}-\vec{R}_1\right)+V_{B_2}\left(\vec{r}-\vec{R}-\vec{R}_2\right)+V_{B_3}\left(\vec{r}-\vec{R}-\vec{R}_3\right). \tag{13}$$

Substituting Eq.(13) into Eq.(3), considering wave function $\varphi(\vec{r})=\sqrt{\lambda^5/\pi}r\cos\theta e^{-\lambda r}$ as $\pi$-electron wave function and using the conclusions of Refs.22 and 23,[22-23] we can obtain the analytic expressions of transfer integrals as

$$\gamma_1=s_1\lambda(\lambda-Z)/2+(-Z)\left(J_1+\sum_{n=2}^{3}J_n\right),$$

$$\gamma_2=\gamma_1,$$

$$\gamma_3=s_3\lambda(\lambda-Z)/2+(-Z)J_1'+2(-Z)J_3', \tag{14a}$$

$$J_1=\lambda\left(\frac{1}{2}+\frac{1}{6}\eta_1^2\right)e^{-\eta_1},$$

$$J_n(n=2,3)=\left(\frac{\lambda}{\eta_n}+\frac{\lambda\eta_1}{2\eta_n^2}\right)S+\frac{\lambda e^{-2|\eta_n|}}{\eta_n\eta_1^3}K-\frac{\lambda e^{-2|\eta_n|}}{\eta_n^2\eta_1^4}T+\frac{\lambda\eta_n e^{-2|\eta_n|}}{16\eta_1^4}U+..., \tag{14b}$$

$$S=\left(1+\eta_1+\frac{2}{5}\eta_1^2+\frac{1}{15}\eta_1^3\right)e^{-\eta_1},$$

$$K=\left(\eta_n^3+9\eta_n^2+\frac{63}{2}\eta_n+\eta_n\eta_1^2+2\eta_1^2+42\right)\left(\sinh\eta_1-\eta_1\cosh\eta_1\right)$$
$$+\left(2\eta_1^2\eta_n^2+9\eta_1^2\eta_n+13\eta_1^2\right)\sinh\eta_1,$$

$$T=A_1F_5+A_2F_4+A_3F_3+A_4F_2+A_5F_1,$$

$$U=2A_1F_2+2A_2F_1+A_3+A_4\left(\frac{e^{-2|\eta_n|}}{\eta_n}-E(\eta_n)\right), \tag{14c}$$

$$A_1=\left(2\eta_1^2-6\eta_1+6\right)e^{\eta_1}-\left(2\eta_1^2+6\eta_1+6\right)e^{-\eta_1},$$

$$A_2 = \left(-4\eta_1^3 + 24\eta_1^2 - 60\eta_1 + 60\right)e^{\eta_1} - \left(4\eta_1^3 + 24\eta_1^2 + 60\eta_1 + 60\right)e^{-\eta_1},$$

$$A_3 = \left(2\eta_1^4 - 24\eta_1^3 + 114\eta_1^2 - 270\eta_1 + 270\right)e^{\eta_1}$$
$$-\left(2\eta_1^4 + 24\eta_1^3 + 114\eta_1^2 + 270\eta_1 + 270\right)e^{-\eta_1},$$

$$A_4 = \left(6\eta_1^4 - 60\eta_1^3 + 270\eta_1^2 - 630\eta_1 + 630\right)e^{\eta_1}$$
$$-\left(6\eta_1^4 + 60\eta_1^3 + 270\eta_1^2 + 630\eta_1 + 630\right)e^{-\eta_1},$$

$$A_5 = A_4, \tag{14d}$$

$$F_1 = \frac{1}{2}\eta_n + \frac{1}{4},$$
$$F_2 = \frac{1}{2}\eta_n^2 + \frac{1}{2}\eta_n + \frac{1}{4},$$
$$F_3 = \frac{1}{2}\eta_n^3 + \frac{3}{4}\eta_n^2 + \frac{3}{4}\eta_n + \frac{3}{8},$$
$$F_4 = \frac{1}{2}\eta_n^4 + \eta_n^3 + \frac{3}{2}\eta_n^2 + \frac{3}{2}\eta_n + \frac{3}{4},$$
$$F_5 = \frac{1}{2}\eta_n^5 + \frac{5}{4}\eta_n^4 + \frac{5}{2}\eta_n^3 + \frac{15}{4}\eta_n^2 + \frac{15}{4}\eta_n + \frac{15}{8},$$

$$E(\eta_n) = \int_{2\eta_n}^{\infty} \frac{1}{t}e^{-t}dt, \tag{14e}$$

$$\eta_1 = \lambda R_1, \quad \eta_n = \lambda R_n, \quad (n = 2,3), \tag{14f}$$

where $J_1'$ is equal to $J_1$ that the $\eta_1$ is displaced by $\lambda R_3$, $J_3'$ is equal to $J_3$ that the $\eta_1$ is also displaced by $\lambda R_3$ and $\eta_3$ is displaced by $\lambda R_1$.

According to the conclusions of Refs.22 and 23, we can also derive the overlap integrals and $\varepsilon_{2p}$ as

$$\varepsilon_{2p} = \lambda(\lambda - Z)/2 - Z\sum_{i=1}^{3} I_i,$$

$$I_i = \frac{1}{R_i} - \frac{3}{2\lambda^2 R_i^3} + \left(\frac{\lambda}{2} + \frac{2}{R_i} + \frac{3}{\lambda R_i^2} + \frac{3}{2\lambda^2 R_i^3}\right)e^{-2\lambda R_i},$$

$$s_i = \left(1 + \lambda R_i + \frac{2\lambda^2 R_i^2}{5} + \frac{\lambda^3 R_i^3}{15}\right)e^{-\lambda R_i} \qquad (i = 1,2,3), \tag{15}$$

where $R_1$, $R_2$, $R_3$ are given by Eq. (11), but the $R_1$, $R_2$, $R_3$ of Eqs. (14) and (15) are all dimensionless that have been divided by the Bohr radius $a_0$, $\lambda$ and $Z$ are the Slater orbital parameter and the effective nuclear charge number of carbon atoms. Based on above equations, we have established the analytic relations between the energy band of $\pi$-electrons in ZGR and the tensile force.

In fact, above equations also give the analytic relations between the energy band of

$\pi$ -electrons in AGR and the tensile force if the Eq. (11) is modified by

$$R_1 = \frac{a_{c-c}}{2\cos\left(\frac{\pi}{3}-\frac{1}{2}\left[(\sqrt{3}+2f_N)-\sqrt{(\sqrt{3}+2f_N)^2+8f_N/\sqrt{3}}\right]\right)},$$

$$R_2 = R_1, \quad R_3 = a_{c-c},$$

$$\alpha = \frac{\pi}{3}-\frac{1}{2}(\sqrt{3}+2f_N)+\frac{1}{2}\sqrt{(\sqrt{3}+2f_N)^2+8f_N/\sqrt{3}}. \tag{16}$$

For showing clearly the changeable relations of ZGR and AGR as the tensile force, Figure.2 is given to show the numerical relations of half-bond angle $\alpha$ ,bond-lengths $R_1(=R_2)$, $R_3$ and transfer integrals $\gamma_1(=\gamma_2), \gamma_3$ with the tensile force which acts on the zigzag edges of graphene ribbon and the armchair edges of graphene ribbon. Here the dimensionless value $f_N = f/(k_0 a_{c-c})$ is taken as 0~0.1.

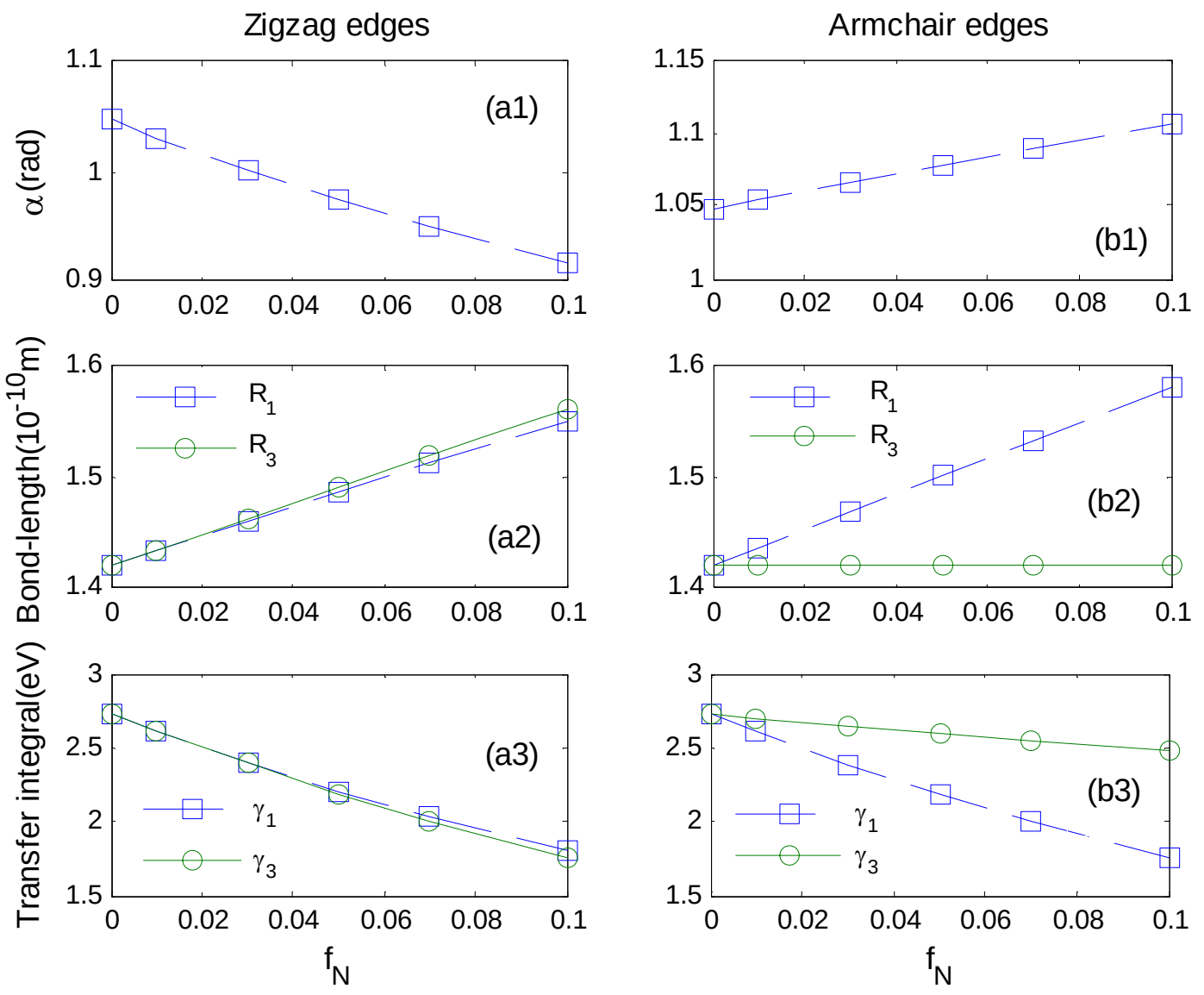

**FIG. 2.** (Color online) The relation graphs of half-bond angle $\alpha$ ,bond-lengths $R_1(=R_2)$, $R_3$ and transfer integrals $\gamma_1(=\gamma_2), \gamma_3$ with the tensile force which acts on the zigzag edges of graphene ribbon (three graphs in left) and the armchair edges of graphene ribbon (three graphs in right). The data values of Figs. (a1), (a2), (b1) and (b2) are obtained by Eqs. (11) and (16), the data values of Figs.(a3) and (b3) are derived by Eqs. (11), (14), (15) and (16) based on effective nuclear charge number $Z$ =2.26, Slater orbital parameter $\lambda$ =2.18.

In Figs.2 (a1) and 2(b1), the increasing or decreasing of half-bond angle $\alpha$ is obvious according to Eqs.(11) and (16) because of the tensile force. In Fig.2 (a2), the increasing of bond-lengths $R_1(=R_2)$, $R_3$ is also obvious due to the force, and the values of $R_1$ and $R_3$ are closed because $f_N$ is very small and the edge effect of graphene ribbon is neglected, which are also the reason of the bond-length $R_3$ for AGR is not changeable in Fig.2(b2), these phenomena are reasonable according to the mechanical knowledge. In Figs.2(a3) and 2(b3), the data relations of transfer integrals $\gamma_1$ ( $=\gamma_2$ ), $\gamma_3$ for ZGR and AGR with the tensile force are obtained by Eqs. (11), (14), (15) and (16) based on effective nuclear charge number $Z$ =2.26, Slater orbital parameter $\lambda$ =2.18. These values of $Z$ and $\lambda$ are taken for getting the transfer integral without tensile force $\gamma_0$ =-2.734eV and the overlap integral without tensile force $s_0$ =0.098, because $\gamma_0$ is between -2.5 and -3eV and $s_0$ is below 0.1for fitting the experimental or first-principles data.[14] In Fig. 2(a3), the changing of $\gamma_1$ is closed to $\gamma_3$ came from the changing values of $R_1$ and $R_3$ are closed because the transfer integral is related to the bond-length. In Fig. 2(b3), the decreasing of $\gamma_3$ is small, which can be explained that the Hamiltonian $H$ is changeable (because the Hamiltonian $H$ changes as the tensile force changes) although $R_3$ is a constant.

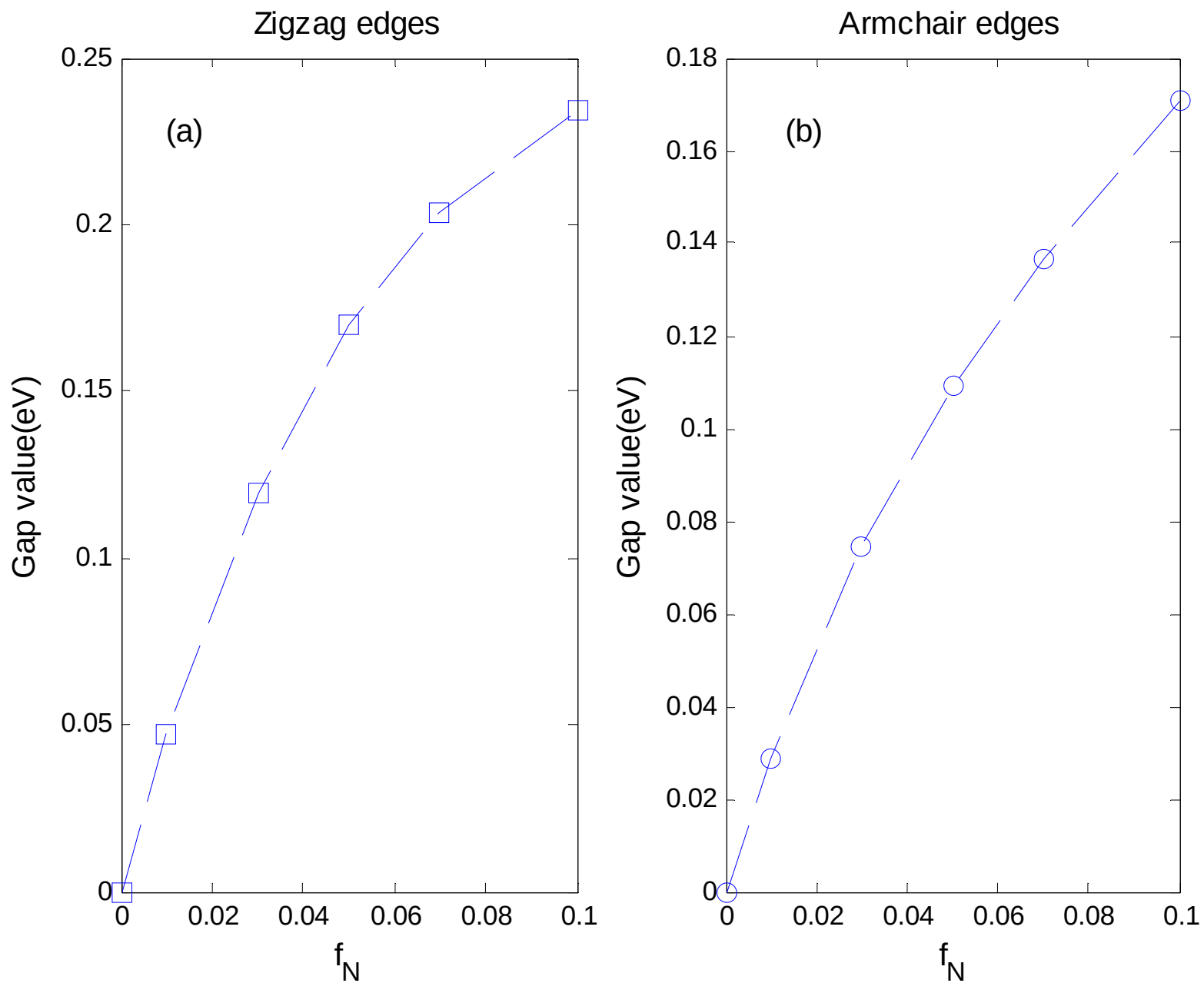

**FIG. 3.** (Color online) The relation graphs between the energy gap values and the tensile force at $K$ point. The left graph shows the gap values of the graphene ribbon with zigzag edges drawn by the tensile force, the right graph shows the gap values of the graphene ribbon with armchair edges drawn by the tensile force. These data values are derived by the Eq. (12), and above analytic relations between the energy bands of $\pi$ -electrons in ZGR, AGR and the tensile force.

In many literatures, $\varepsilon_{2p}$ is considered as a parameter which is often taken as zero or closed to zero,[13,14] in fact, the $\varepsilon_{2p}=\left\langle\varphi_A\left(\vec{r}-\vec{R}\right)\middle|H\middle|\varphi_A\left(\vec{r}-\vec{R}\right)\right\rangle$ is much larger than zero whether the ZGR and AGR are drawn by the force or not though the calculation of above formulas based on $\pi$ -electron wave function $\varphi(\vec{r})=\sqrt{\lambda^5/\pi}r\cos\theta e^{-\lambda r}$ and $Z$ =2.26, $\lambda$ =2.18, so the energy values $E^{\pm}$ of graphene ribbons are dependent mainly from $\varepsilon_{2p}$ besides other parameters are related in Eq. (5), which produces that the $E^{\pm}$ of graphene ribbons without tensile force are not closed to zero at $K$ point due to $\varepsilon_{2p}$ is not closed to zero, but this defect can be neglected in calculating energy difference $\Delta E$ between two energy bands in all Brillouin zone because $\Delta E$ is independent approximately from $\varepsilon_{2p}$ according to Eq. (5) . So we can obtain the accurate data values of $\Delta E$ (including the gap values at $K$ or $K'$ point) based on Eq. (5) because $\varepsilon_{2p}$ can be neglected and $\gamma_0$, $s_0$ are taken for fitting the experimental or first-principles data though $Z$ =2.26, $\lambda$ =2.18. In Fig. 3, we show the data relations between the gap values of ZGR (shown in Fig. 3(a)), AGR (shown in Fig. 3(b)) and the tensile force at the $K$ point according to the Eq. (12), and above analytic relations between the energy bands of $\pi$-electrons in ZGR, AGR and the tensile force based on $Z$ =2.26, $\lambda$ =2.18. In this figure, we find that the tensile force can open an energy gap of ZGR or AGR at the $K$ point and the gap values of ZGR and AGR both increase as the force increases, moreover the gap values of ZGR increase more far than the AGR at the $K$ point as the tensile force increases. For these important phenomenon, we can explain that the opening of energy gap for graphene ribbons is due to the deformation of graphene ribbons produced from the force, and the deformation that zigzag edges are drawn by the fore gives an energy gap more easily.

In conclusion, we have investigated the electronic structures of ZGR and AGR drawn by the tensile force according to the tight-binding approximation, and obtained the analytic relations between the energy bands of $\pi$-electrons in ZGR, AGR and the tensile force based on only considering the nearest-neighbor interaction and wave function $\varphi(\vec{r}) = \sqrt{\lambda^5 / \pi} r \cos\theta e^{-\lambda r}$ is regarded as $\pi$-electron wave function. Though the numerical calculation , we give the numerical relations of half-bond angles, bond-lengths, transfer integrals and the gap values at $K$ point with the tensile force for ZGR and AGR, and find that the tensile force can open an energy gap of ZGR or AGR at the $K$ point and the tensile force perpendicular to the zigzag edges can open energy gap more easily besides two gap values of ZGR and AGR at the $K$ point both increase as the tensile force increases.

---

* tgp6463@zjnu.cn